\documentclass[conference]{IEEEtran}
\IEEEoverridecommandlockouts

\usepackage{cite}
\usepackage{amsmath,amssymb,amsfonts}
\usepackage{algorithm}
\usepackage{algorithmic}
\usepackage{graphicx}
\usepackage{textcomp}
\usepackage{xcolor}
\usepackage{bm}
\usepackage{fancyhdr}

\fancypagestyle{firstpage}{
  \fancyhf{}
  \fancyhead[L]{PREPRINT}
  
}

\def\BibTeX{{\rm B\kern-.05em{\sc i\kern-.025em b}\kern-.08em
    T\kern-.1667em\lower.7ex\hbox{E}\kern-.125emX}}
\begin{document}

\title{External Sinkhole Attack Detection in Large-Scale WSNs Using Metaheuristic Feature Selection\\
\thanks{This work was supported by JST BOOST, Japan Grant Number \mbox{JPMJBS2420.} (Corresponding author: Seungwoo Han, han@sip.tuat.ac.jp.)

© 2026 IEEE. Personal use of this material is permitted. Permission from
IEEE must be obtained for all other uses, in any current or future media,
including reprinting/republishing this material for advertising or promotional
purposes, creating new collective works, for resale or redistribution to servers
or lists, or reuse of any copyrighted component of this work in other works.

}
}

\author{
\IEEEauthorblockN{
$^1$Seungwoo Han,
$^1$Sawako Kitagata, 
$^1$Ingon Chanpornpakdi, 
$^1$Toshihisa Tanaka,
and
$^2$Su Man Nam
}
\IEEEauthorblockA{
$^{1}$\textit{Department of Electrical Engineering and Computer Science, Tokyo University of Agriculture and Technology}, Tokyo, Japan \\
$^{2}$\textit{Department of Digital Security, Cheongju University}, Cheongju-si, Republic of Korea
}
han@sip.tuat.ac.jp, smnam@cju.ac.kr
}

\maketitle
\thispagestyle{firstpage}

\begin{abstract}
Sinkhole attacks in large-scale wireless sensor networks (WSNs) pose a serious threat to network functionality. This paper presents a metaheuristic feature selection for sinkhole attack detection using the bee swarm optimization (BSO) algorithm. In an external sinkhole attack simulation with 2000 nodes deployed over a 3000 $\times$ 3000 m$^2$ field, the proposed method achieves a detection accuracy of 0.997 while reducing the 16-feature set to eight features.
\end{abstract}


\section{Introduction}
Large-scale wireless sensor networks (WSNs) comprise hundreds to thousands of sensor nodes and enable wide-area monitoring through communication with a base station \cite{Nam2021, Han2026}. 
WSNs are vulnerable to internal and external sinkhole attacks \cite{Shafiei2014} due to their wireless characteristics. An internal sinkhole attack compromises normal nodes by installing malicious software. On the contrary, an external attack introduces new malicious nodes into the network from outside \cite{Shafiei2014}. 

Machine learning (ML)-based approaches are  powerful solutions for automatic sinkhole attack detection \cite{Ioannou2020, Aissaoui2021, Hasan2025, Bensaid2025, Khedr2024, Han2026}. However, some of these studies are 
limited to small-scale environments \cite{Ioannou2020} and internal attack scenarios \cite{Ioannou2020, Aissaoui2021, Hasan2025, Bensaid2025, Khedr2024}, or focus on employing a large number of features \cite{Han2026}. In particular, since WSNs operate under resource constraints, low-dimensional feature inputs are essential for reliable operation. Nevertheless, many prior studies have left external attack detection and feature reduction insufficiently explored.

In this paper, we aim to detect external sinkhole attacks more efficiently by employing a metaheuristic feature search-based intrusion detection system using the bee swarm optimization (BSO) \cite{Drias2005, Sadeg2015}, assuming a deployment scenario in which the trained inference model operates on the selected features. BSO is well suited for this task due to its ability to explore large search spaces while avoiding local optima. We not only validate our proposed approach in a large-scale environment but also compare it with various feature selection methods. Moreover, we measure the floating-point operations per-sample (FLOPs/sample) during inference to quantify the reduction in computational cost, verifying that the selected features enable more efficient inference.

\section{Method}
\subsection{Data Generation and Preprocessing}
We conducted an external sinkhole attack simulation using PyWSNSim \cite{Han2026} with the following configuration: a 2D field size of 3000 $\times$ 3000 m$^2$, 2000 MICAz \cite{Memsic2003micaz} sensor nodes, a base station located at (2000, 2000), the Dijkstra routing protocol, 4 attacker nodes, an attack probability of 90$\%$, and an attack range of 150 m. The final feature set is composed of 16 attributes, as summarized in Table \ref{tab:features}. For a description of each attribute, please refer to our previous study and PyWSNSim source code \cite{Han2026, PyWSNSimgit}. Among the features, the \texttt{hop\_count} column contained infinity values. We replaced infinity values with the maximum finite value in that column and added one to handle an arithmetic error. The simulation data were randomly partitioned such that 80$\%$ and 20$\%$ were allocated for ML model training and testing, and the features were then z-normalized as follows:

\begin{equation}
z = \frac{x - \mu}{\sigma}
\label{eq:normalization}
\end{equation}

where $x$ denotes the feature value, $\mu$ the mean of the feature, and $\sigma$ the standard deviation. The training set contains 1559 normal nodes and 41 affected nodes, and the test set contains 390 normal nodes and 10 affected nodes.

\begin{table*}[t]
\centering
\caption{Feature List from PyWSNSim-Based Sinkhole Attack Simulation.}
\label{tab:features}
\begin{tabular}{llll}
\hline
\multicolumn{4}{c}{\textbf{Feature list from PyWSNSim \cite{Han2026}}} \\
\hline
0. \texttt{pos\_x}                 & 1. \texttt{pos\_y}          & 2. 
\texttt{hop\_count}          & 3. \texttt{neighbor\_nodes}   \\
4. \texttt{route\_changes}  & 5. \texttt{distance\_to\_bs}        & 6. \texttt{tx\_count}        & 7. \texttt{rx\_count}  \\
8. \texttt{energy\_level}             & 9. \texttt{initial\_energy}               & 10. \texttt{energy\_percentage}      & 11. \texttt{consumed\_energy\_tx}   \\ 12.
\texttt{consumed\_energy\_rx}        & 13. \texttt{total\_consumed\_energy}               & 14. \texttt{tx\_energy\_per\_byte}    & 15. \texttt{rx\_energy\_per\_byte}   \\
\hline
\end{tabular}
\end{table*}

\begin{algorithm}
\caption{BSO-based feature selection \cite{Sadeg2015}}
\label{algo:BSO-FS}
\begin{algorithmic}[1]
    \STATE \textbf{Parameters} $MaxChances \leftarrow 3;$ $NumBees \leftarrow 10;$ $MaxIter \leftarrow 5$; $LocalIter \leftarrow 5$; $flip \leftarrow 4$;
    $Fitness Function \ f(\cdot) \leftarrow Acc.$;

    $TabuList \leftarrow \emptyset$; $ResList \leftarrow \emptyset$;
    \STATE $RefF \leftarrow$ randomly choose sinkhole attack features; 
    $BestF \leftarrow RefF$;
    \WHILE{number of iterations $< MaxIter$}
        \STATE Insert $RefF$ into $TabuList$
        \FOR{each bee $k = 1, \ldots, NumBees$}
            \STATE Perform local search iterations $LocalIter$ via Eq.~\eqref{eq:flip}
            
            store result in $ResList$
        \ENDFOR
        \IF{$MaxChances > 0$}
            \STATE $RefF \leftarrow \arg\max_{p \in ResList}  f(p)$ via ML model
            \IF{$f(RefF) > f(BestF)$}
                \STATE $BestF \leftarrow RefF$; 
            \ELSE
                \STATE $MaxChances \leftarrow MaxChances - 1$
            \ENDIF
        \ELSE
           \STATE $RefF \leftarrow \arg\max_{p \in ResList} \min_{q \in TabuList} D(p, q)$;
           where $D(p, q) = \sum_{i=1}^{N_f} |p_i - q_i|$
        \ENDIF
    \ENDWHILE
    \RETURN $BestF$
\end{algorithmic}
\end{algorithm}

\subsection{BSO-based Feature Selection}

BSO \cite{Drias2005} is a swarm optimization algorithm inspired by the behavior of bees. The pseudo-code and parameters of the BSO-based feature selection \cite{Sadeg2015} are presented in Algorithm~\ref{algo:BSO-FS}, with its specific parameter values determined through our empirical selection. To generate neighboring solutions, each bee flips a bit in the current feature vector $v$ according to Eq.~\eqref{eq:flip}:

\begin{equation}
v_j \leftarrow (v_j + 1) \bmod 2
\label{eq:flip}
\end{equation}

where $j$ is determined as follows:

\begin{equation}
\scalebox{0.8}{$
j = \begin{cases}
    \text{flip} \cdot l + (k - 1) & \text{if } k \le \text{flip} \\
    \lfloor N_{f}/\text{flip} \rfloor \cdot (k - \text{flip} - 1) + l & 
    \begin{aligned}[t]
    \text{if } &\text{flip} < k \le 2 \cdot \text{flip}, \\
    &l < \lfloor N_{f}/\text{flip} \rfloor, \text{ and} \\
    &\text{flip} \cdot l + (k - \text{flip} - 1) < N_{f}
    \end{aligned} \\
    r \sim \mathcal{U}\{0, N_{f}-1\} & \text{otherwise}
\end{cases}$}
\label{eq:j}
\end{equation}

where $k$ is the bee index, $l$ is the loop index, and $N_{f}$ is the total number of features. $r\sim \mathcal{U}$ is random sampling.

The ML model used to measure detection accuracy (Acc.) was radial basis function-based support vector machine (SVM) \cite{Cortes1995}, and the default hyper-parameters provided by scikit-learn 1.8.0 \cite{JMLR:v12:pedregosa11a} were used.

\section{Results}
\subsection{Evaluation of Feature Selection Performance}

Table \ref{tab:ablation} shows the inference performance --- accuracy, precision (Pre.), recall (Rec.), and FLOPs/sample --- for all features and the three feature selection methods. Accuracy, precision, recall, and FLOPs/sample are defined as follows:

\begin{equation}
\text{Accuracy} = \frac{TP + TN}{TP + TN + FP + FN}
\label{eq:acc}
\end{equation}

\begin{equation}
\text{Precision} = \frac{TP}{TP + FP}
\label{eq:precision}
\end{equation}

\begin{equation}
\text{Recall} = \frac{TP}{TP + FN}
\label{eq:recall}
\end{equation}

\begin{equation}
\text{FLOPs/sample}_{\text{SVM\_inference}} = 2 \cdot N_{sv} \cdot N_{f}
\label{eq:flops}
\end{equation}

where $TP$, $TN$, $FP$, $FN$, $N_{sv}$ denote number of true positives, true negatives, false positives, false negatives, and support vectors, respectively. The FLOPs were measured using the FLOPpy \cite{Scala2026}. 

All features achieved an accuracy of 0.995. Low-variance threshold elimination retained 13 features and achieved the same accuracy, while least absolute shrinkage and selection operator (LASSO) \cite{Tibshirani1996} reduced the feature count to eight features with an accuracy of 0.992. Notably, the BSO-based method achieved the highest accuracy of 0.997, using eight features with indices 0, 1, 2, 5, 6, 13, 14, and 15. 

Beyond accuracy, the proposed method also attained the highest precision of 1.000 and recall of 0.900, whereas LASSO, despite using the same number of features, yielded a lower precision of 0.888 and recall of 0.800. In terms of computational cost, the proposed method required only 912 FLOPs/sample, the lowest among all methods and roughly a 50$\%$ reduction compared with using all features. These results suggest that the BSO-based method achieves accuracy on par with the other approaches while attaining the best overall performance across the three metrics at the lowest computational cost.

\begin{table}[t]
\centering
\caption{Performance comparison of different feature selection methods.}
\label{tab:ablation}
\scalebox{0.9}{
\begin{tabular}{lccccc}
\hline
\textbf{Method} & \textbf{$\bm{N_f}$} & \textbf{Acc.} & \textbf{Pre.} &  \textbf{Rec.} & \textbf{FLOPs/sample}\\
\hline
All features (Baseline) & 16 & 0.995 & 1.000 & 0.800 & 1952 \\
Low-variance threshold & 13 & 0.995 & 1.000 & 0.800 & 1586 \\
LASSO & 8 & 0.992 & 0.888 & 0.800 & 1040 \\
\textbf{BSO (Proposed)} & \textbf{8}  & \textbf{0.997} & \textbf{1.000} & \textbf{0.900} & \textbf{912}\\
\hline
\end{tabular}
}
\end{table}

\subsection{Comparison of Existing Works}

\begin{table*}[t]
\centering
\caption{Comparison of experiment scale and detection performance with existing studies. Values denote figures reported in the referenced studies.}
\label{tab:acc_result}
\begin{tabular}{lccccccccc}
\hline
\textbf{Author} & \textbf{Inference model} & \textbf{Attack type} & \textbf{Field size}& \textbf{Node count} & \textbf{$\bm{N_f}$} & \textbf{Acc.} & \textbf{Pre.} & \textbf{Rec.} & \textbf{FLOPs/sample}\\
\hline
Ioannou \textit{et al.} \cite{Ioannou2020} & Logistic regression
& Inside & 5 $\times$ 5 & 25 & 4 & 1.000 & 1.000 & 1.000 & N.R. \\
Aissaoui \textit{et al.} \cite{Aissaoui2021} & SVM & Inside & 100 $\times$ 100 & 20 & 2 & 1.000 & N.R. & N.R. & N.R. \\
Hasan \textit{et al.} \cite{Hasan2025} & Random forest & Inside & 500 $\times$ 500 & 50 & 5 & 1.000 & 1.000 & 1.000 & N.R. \\
Bensaid \textit{et al.} \cite{Bensaid2025} & Neural networks & Inside & 1000 $\times$ 1000 & 45  & 8 & 0.971 & N.R. & N.R. & N.R.\\
Khedr \textit{et al.} \cite{Khedr2024} & Neural networks & Inside & 1100 $\times$ 1100 & 500 & 4 & 0.954 & 0.972 & 0.978 & N.R.\\
Han \textit{et al.} \cite{Han2026} & SVM & Outside & 2000 $\times$ 2000 & 1000 & 16 & 0.980  & 1.000 & 0.764 & N.R.\\
This study & SVM & Outside & 3000 $\times$ 3000 & 2000 & 8 & 0.997 & 1.000 & 0.900 & 912\\
\hline
\raggedright\footnotesize N.R. : Not reported.
\end{tabular}
\end{table*}

Table \ref{tab:acc_result} compares our result with existing ML-based sinkhole attack detection studies. The results demonstrate that the proposed method maintains high accuracy even in a large-scale environment containing a large number of nodes. Moreover, unlike some existing studies that report only accuracy and omit precision and recall, our study evaluates the model comprehensively across precision and recall as well as FLOPs/sample.

\section{Discussion}
This study has two points that warrant further validation. First, the performance improvement over the baseline is relatively marginal, and the evaluation is confined to simulations employing an SVM. In addition, the model was evaluated on a random data split, and we plan to conduct a cross-validation in future work given the class imbalance. Second, as BSO is a metaheuristic that necessitates iterative evaluations, its overall computational cost is likely to be substantially higher than that of alternative feature selection methods. However, this cost occurs only once during training, while our efficiency gains apply to the recurring per-sample inference at deployment. Nevertheless, the selection cost should still be weighed against model inference benefit when determining an appropriate feature selection strategy.

\section{Conclusion}
In this paper, we evaluated the effectiveness of metaheuristic feature selection for external sinkhole 
attack detection in large-scale WSNs. The evaluation demonstrates that the feature set can be successfully reduced to a smaller number of features while maintaining a high detection accuracy. Future work will extend this study by designing unsupervised models that achieve both low inference energy consumption and high detection accuracy.

\bibliography{references}
\bibliographystyle{ieeetr}

\end{document}